\documentclass[conference]{IEEEtran}
\usepackage{hyperref}
\usepackage[noadjust]{cite}
\usepackage[cmex10]{amsmath}
\usepackage{esint}
\usepackage{commath}
\usepackage[caption=false,font=footnotesize]{subfig}
\usepackage{url}
\usepackage{graphicx}
\usepackage{amssymb}
\everymath{\displaystyle}
\usepackage{comment}

\usepackage[shortlabels]{enumitem}
\usepackage{nameref}
\usepackage{slashbox}
\usepackage{textcomp}
\usepackage{gensymb}
\usepackage{tikz}
\usetikzlibrary{patterns}
\usepackage{tkz-euclide}

\newcommand{\myexp}{\mathsf{e}}

\newcommand{\myfunctionw}{\vec{w}}
\newcommand{\myfunctionwi}[2][]{\myfunctionw _{#2} ^{#1}}

\mathchardef\UrlBreakPenalty=10000
\mathchardef\UrlBigBreakPenalty=10000

\onecolumn

\begin{document}
\title
{
  A Turning Band Approach to\\
  Kernel Convolution for Arbitrary Surfaces
}
\author
{
\IEEEauthorblockN{Alexander Gribov}
\IEEEauthorblockA
{
  Environmental Systems Research Institute\\
  380 New York Street\\
  Redlands, CA 92373\\
  E-mail: agribov@esri.com}
}

\maketitle

\begin{abstract}
\boldmath
One of the most efficient ways to produce unconditional simulations is with the spectral method using fast Fourier transform (FFT) \cite{KernelConvolutionFFT}. But this approach is not applicable to arbitrary surfaces because no regular grid exists. However, points on the arbitrary surface can be generated randomly using uniform distribution to replace a regular grid. This paper will describe a nonstationary kernel convolution approach for data on arbitrary surfaces.
\end{abstract}


\section{Introduction}

The approach to producing unconditional simulations for arbitrary surfaces (sphere, ellipsoid, geoid, torus, etc.) will be described in this paper. The approach is also applicable to nonsmooth surfaces. This paper is based on ideas from a turning band approach (\cite{TwoDimensionalSimulationByTurningBands}, \cite{TurningBandTwoDimensional}, \cite{ClosedFormSolutionsOfTheTwoDimensionalTurningBandEquation}, and \cite{TurningBandSimulationsInThreeDimensions}) and kernel convolution (\cite{KernelConvolution1996}, \cite{KernelConvolution1998}, and \cite{KernelConvolutionFFT}). Similar to the turning band, all processing is done on a set of random lines. The difference is in using a random set of points covering an area of interest and projecting them onto the set of random lines. For example, the set of uniformly distributed points on the surface of a sphere is projected onto the set of random lines. The kernel convolution is applied to each line, and the reconstruction of the process is performed similarly to the turning band approach. Using different kernels will produce a nonstationary solution.




\section
{
  Kernel Convolution by Turning Band
  \label{sec:KernelConvolutionByTurningBand}
}

Here are the steps of the algorithm:
\begin{enumerate}
  \item \label{en:UniformDistributionOnSurface} From a uniform distribution, simulate $N$ points $\vec{s_i}$ on a surface. \footnote{Replacing uniform distribution on the surface by pseudorandom uniform distribution will produce an even coverage and guarantee the stability of the algorithm.}
  \item From a standard normal distribution, simulate value $\xi_i$ for each point.
  \item Simulate $N_d$ random or pseudorandom directions (the algorithm to simulate pseudorandom directions can be found in \cite{RandomDirections} and \cite{TurningBandSimulationsInThreeDimensions}).
  \item Cycle over all directions.
  \begin{enumerate}
    \item Project points to a direction.
    \item Apply kernel convolution (this is usually done using FFT).
  \end{enumerate}
  \item For each location where the result is desired, project that location to all directions and average the results of kernel convolutions.
\end{enumerate}

The applied one-dimensional kernel $k_1{\left( \bullet \right)}$ will correspond to a three-dimensional kernel $k_3{\left( \bullet \right)}$ through integral
\begin{equation}
  k_3{\left( \vec{h} \right)}
  =
  \oiint
  {
    k_1{\left( \left( \vec{h}, \hat{n} \right) \right)}
    \dif{\hat{n}}
  }
  ,
  \label{eq:IntegralKernel}
\end{equation}
where $\left( \vec{x}, \vec{y} \right)$ is a scalar product.

The kernel distance is the direct distance in three-dimensional space.

This is similar to the turning band method, and the relationship between three-dimensional or two-dimensional covariance with one-dimensional covariance can be found in \cite{TwoDimensionalSimulationByTurningBands}, \cite{TurningBandTwoDimensional}, and \cite{ClosedFormSolutionsOfTheTwoDimensionalTurningBandEquation}. The same relationship holds true for the kernel functions. In a three-dimensional case, the formula is
\begin{equation}
  k_1(h) = (h \cdot k_3{\left( h \right)})'
  .
  \label{eq:RelationshipBetweenKernelsThree}
\end{equation}

In two dimensions for kernel $k_2{\left( \bullet \right)}$
\begin{equation*}
  k_1(h)
  =
  k_2{\left( 0 \right)}
  +
  h
  \int\limits_{0}^{h}
  {
    \dfrac
    {
      k_2'{\left( x \right)}
    }
    {
      \left(
        h^2 - x^2
      \right)
      ^
      {
        \frac{1}{2}
      }
    }
    \dif{x}
  }
  =
  k_2{\left( 0 \right)}
  +
  h
  \int\limits_{0}^{\frac{\pi}{2}}
  {
    k_2'
    {
      \left(
        h
        \sin{\left( \Theta \right)}
      \right)
    }
    \dif{\Theta}
  }
  .
\end{equation*}

\section{Kernel Convolution for Unconditional Simulation}

Applying the algorithm from the section ``\nameref{sec:KernelConvolutionByTurningBand}'' for any location $\vec{s}$ will produce a random field approximately equal to
\begin{equation}
  \sum_{i = 0}^{N}
  {
    \left(
      k_3
      {
        \left( \vec{s_i} - \vec{s} \right)
      }
      \cdot
      \xi_i
    \right)
  }
  .
  \label{eq:KernelApplied}
\end{equation}

The approximation is due to the finite number of directions.

The variance at location $\vec{s}$ is approximately equal to
\begin{equation}
  \sum_{i = 0}^{N}
  {
    k_3
    ^2
    {
      \left( \vec{s_i} - \vec{s} \right)
    }
  }
  .
  \label{eq:KernelVariance}
\end{equation}

For the sphere, the estimate of \eqref{eq:KernelVariance} does not depend on location $\vec{s}$. This does not hold true for other surfaces. However, for any surface with finite sampling $N$, it is not a constant. Calculating this value is important in standardizing unconditional simulations. The estimate of \eqref{eq:KernelVariance} can be found by applying the algorithm from the section ``\nameref{sec:KernelConvolutionByTurningBand}'' for the square of the kernel for the same set of points with values of one. Dividing the approximation of \eqref{eq:KernelApplied} by the square root of the approximation of \eqref{eq:KernelVariance} will produce an approximation of the Gaussian process with unit variance.

Another way to estimate \eqref{eq:KernelVariance} is by producing many unconditional simulations and estimating the variance from them. This will also allow the estimation of covariances.

Compared to the turning band approach, using this algorithm is significantly slower. The slowest part is projecting simulated points to directions. Notice that the simulated points do not depend on the kernel. Therefore, to speed up the approach, it is possible to simulate points and project them to directions in advance. Combining them by bins will significantly reduce the amount of memory necessary to store precalculated results.

\section{Conditional Simulations}

To perform conditional simulations, it is necessary to know the covariance function. For a sphere, the covariance function is isotropic and does not depend on location. Therefore, it can be derived directly from the kernel as described in \cite{CompactCovarianceModelOnASphere} and \cite{KernelConvolutionForRingsOnPlane}. For arbitrary surfaces, it is possible to make many unconditional simulations and from them estimate the covariance function. Conditional simulations are obtainable from unconditional simulations through solutions of the Kriging system \cite{ConditionalSimulationsGeostatisticsForConditionalSimulationOfOreBodies}, \cite{ConditionalSimulationsBook}, and \cite{ConditionalSimulationsSpatialPredictionSpatialSamplingAndMeasurementError}. \cite{GeostatisticalMappingWithContinuousMovingNeighborhood} describes the approach to construct conditional simulations without abrupt changes.

\section{Class of Exponential Kernels}

Let the kernel function have the form
\begin{equation}
  k_3{\left( h \right)}
  =
  \sum_{i}
  {
    \left(
      P_i{\left( h \right)}
      \cdot
      \myexp^{-\Theta_i \cdot h}
    \right)
  }
  ,
  \label{eq:Form}
\end{equation}
where $P_i$ is a polynomial.

The result of the convolution can be obtained by the sequential application of the kernel. The technique is described in \cite{Smoothing} and \cite{SmoothingNetwork} with the difference being that the integration is replaced by summation. This reduces the complexity of the kernel convolution from $ O{\left( n \cdot \ln{n} \right)} $ to $ O{\left( n \right)} $, where $n$ is the number of bins on the line.

For finite kernel
\begin{equation}
  k_3{\left( h \right)}
  =
  \left\{
  {
    \begin{aligned}
      &\sum_{i}
      {
        \left(
          P_i{\left( h \right)}
          \cdot
          \myexp^{-\Theta_i \cdot h}
        \right)
      }
      ,
      &\text{ if }
      h < T,\\
      &0,
      &\text{ otherwise},
    \end{aligned}
  }
  \right.
  \label{eq:FormFinite}
\end{equation}
where $T$ is the kernel width. To be able to evaluate $k_1{\left( h \right)}$ from \eqref{eq:RelationshipBetweenKernelsThree}, it is necessary to satisfy $\lim\limits_{h \rightarrow T^{-}}{k_3{\left( h \right)}} = 0$. Otherwise, $k_1{\left( h \right)}$ will not be finite.

There are several advantages of using the finite kernel:
\begin{itemize}
  \item Ability to produce unconditional simulations for a large area by partitioning it into small pieces (tiles) for independent processing.
  \item Ability to have a flexible class of kernels in form \eqref{eq:FormFinite}.
\end{itemize}

The ability to have a flexible class of kernels is important for nonstationary simulations. It allows not only changes in kernel bandwidth but in their shape as well. One way to construct a flexible class of kernels in form \eqref{eq:FormFinite} is by changing the polynomial part. It is reasonable to use decreasing kernels because the majority of physical processes gradually decreases dependence over distance. From~\cite{MonotonicFunction}, a Bernstein density (a weighted sum of beta probability distribution functions defined on the unit interval
$ \left[ 0 , 1 \right] $) is defined as
\newcommand{\myfunctionbeta}[1]
{
  Beta
  {
    \left(
      #1
      \mid
      j , m - j + 1
    \right)
  }
}
\newcommand{\myfunctionbetacorrection}[2][]
{
  \sum _{j = 1} ^{m}
    \left(
      \myfunctionwi[#1]{j}
      \cdot
      \myfunctionbeta{#2}
    \right)
}
\newcommand{\myfunctionbetacorrectionintegrated}[2][]
{
  \sum _{j = 1} ^{m}
    \left(
      \myfunctionwi[#1]{j}
      \int\limits_{x}^{1}
      {
        \myfunctionbeta{#2}
        \dif{y}
      }
    \right)
}
\begin{equation*}
  p{\left( x \mid m , \myfunctionw \right)} = \myfunctionbetacorrection{x} \; , \;
  \myfunctionwi{j} \geq 0 \; , \;
  \sum _{j = 1} ^{m} {\myfunctionwi{j} = 1} \; ,
\end{equation*}
where
$ m $ is the number of distribution functions,
$ \myfunctionw $ are weights,
$ Beta{\left( \bullet \right)}$
is the beta probability density function
\begin{equation*}
  \myfunctionbeta{x}
  =
  \dfrac{m !}{\left( j - 1 \right) ! \cdot \left( m - j \right) !}
  \cdot
  x ^{j - 1}
  \cdot
  \left( 1 - x \right) ^{m - j}
  \; .
\end{equation*}

Forming
\begin{equation*}
  P{\left( x \right)}
  =
  \int\limits_{x}^{1}
  {
    p{\left( x \mid m , \myfunctionw \right)}
  }
  =
  \myfunctionbetacorrectionintegrated{y}
\end{equation*}
and multiplying by $\myexp^{-\frac{h}{\Theta}}$ produces a flexible family of kernels. The choice of $\Theta$ is arbitrary. If the value is too large, the floating point evaluation will lead to a high round off error. If the value is too small, it will not produce good family of kernels. Let's define $\Theta = 3$.

\begin{figure} [htb]
  \centering
  \begin{tabular}{c c}
    \includegraphics[width = 8 cm, keepaspectratio]{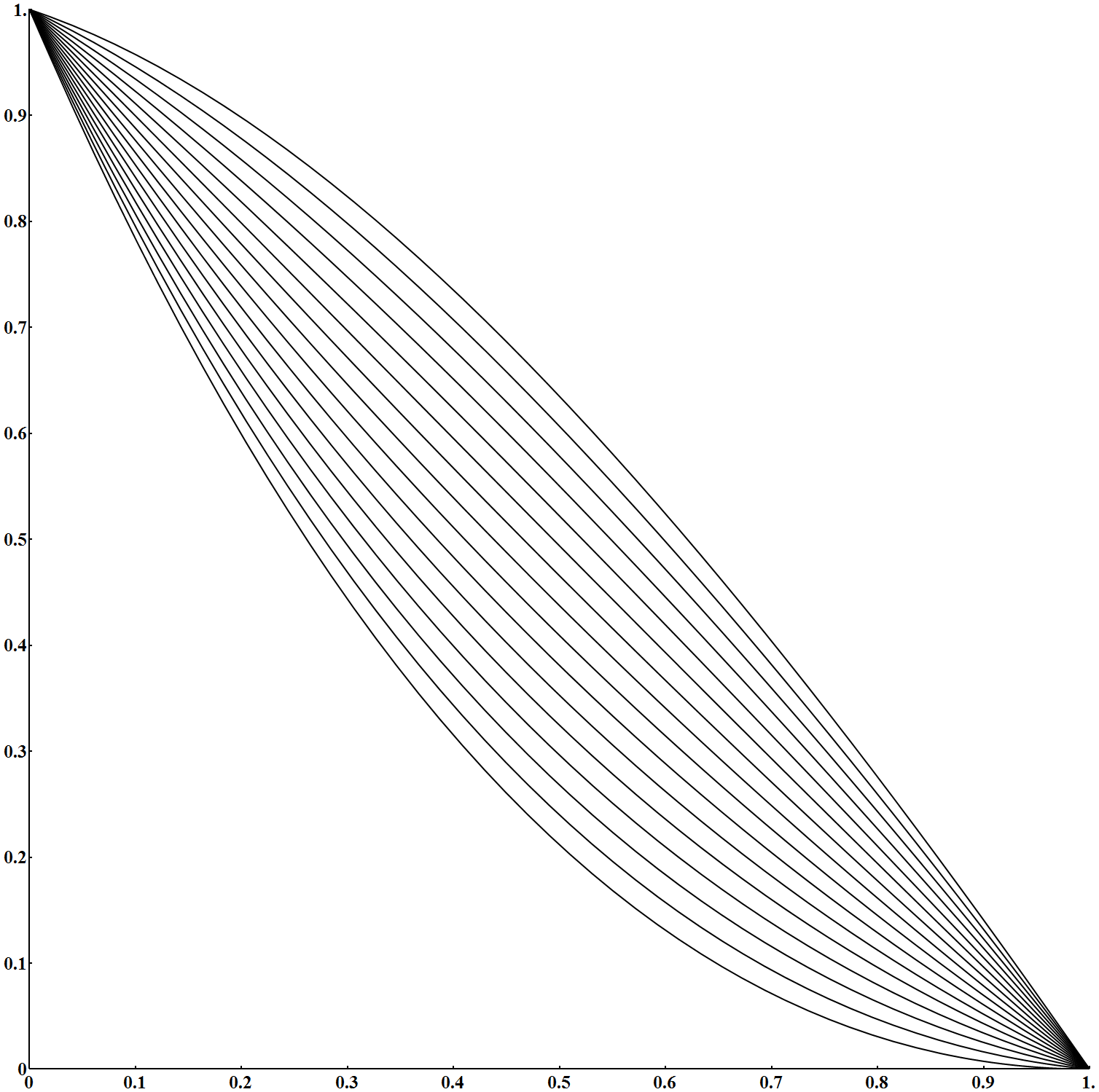} &
    \includegraphics[width = 8 cm, keepaspectratio]{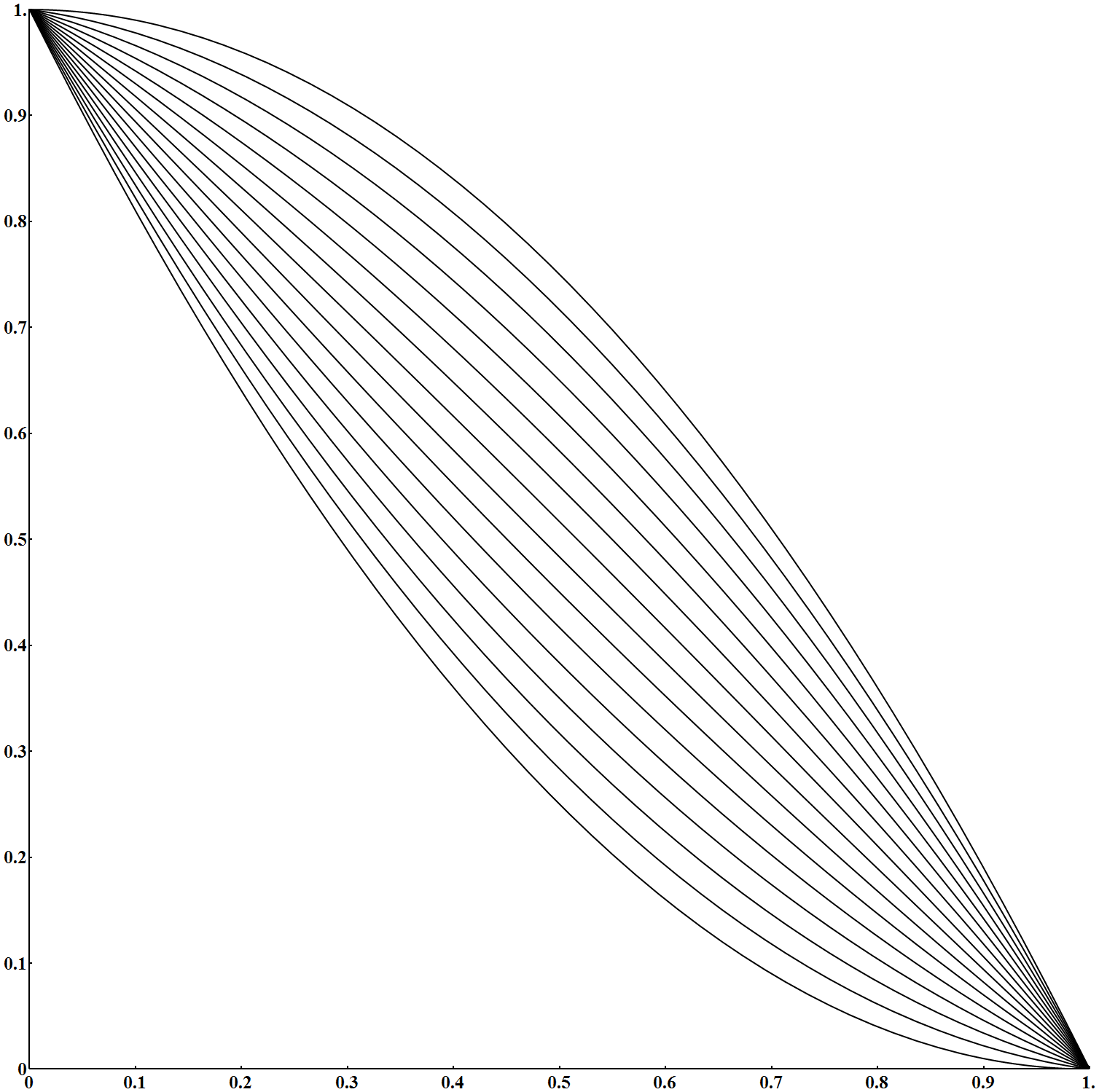} \\
    a) &
    b)
  \end{tabular}
  \caption
  {
    Example of kernels \eqref{eq:KernelsPolynomialExponentTwo}.
    $\myfunctionwi{1} = \dfrac{i}{15}$, $\myfunctionwi{2} = 1 - \myfunctionwi{1}$, $i = \overline{0..15}$. a) With exponent. b) Without exponent.
  }
  \label{fig:KernelsPolynomialExponent}
\end{figure}

In the case of $m = 2$, this becomes
\begin{equation}
  \left\{
    \begin{aligned}
      &
      \left(
        \myfunctionwi{1} \left( 1 - h \right)^2
        +
        \myfunctionwi{2} \left( 1 - h^2 \right)
      \right)
      \cdot
      \myexp^{-\frac{h}{3}}
      ,
      &
      \text{ if }
      h < 1,
      \\
      &
      0
      ,
      &
      \text{otherwise},
    \end{aligned}
  \right.
  \label{eq:KernelsPolynomialExponentTwo}
\end{equation}
where $0 \leq \myfunctionwi{i} \leq 1$, $i = 1, 2$, and $\myfunctionwi{1} + \myfunctionwi{2} = 1$. They are shown in Figure~\ref{fig:KernelsPolynomialExponent}a.

\begin{figure} [htb]
  \centering
  \begin{tabular}{c c}
    \includegraphics[width = 8 cm, keepaspectratio]{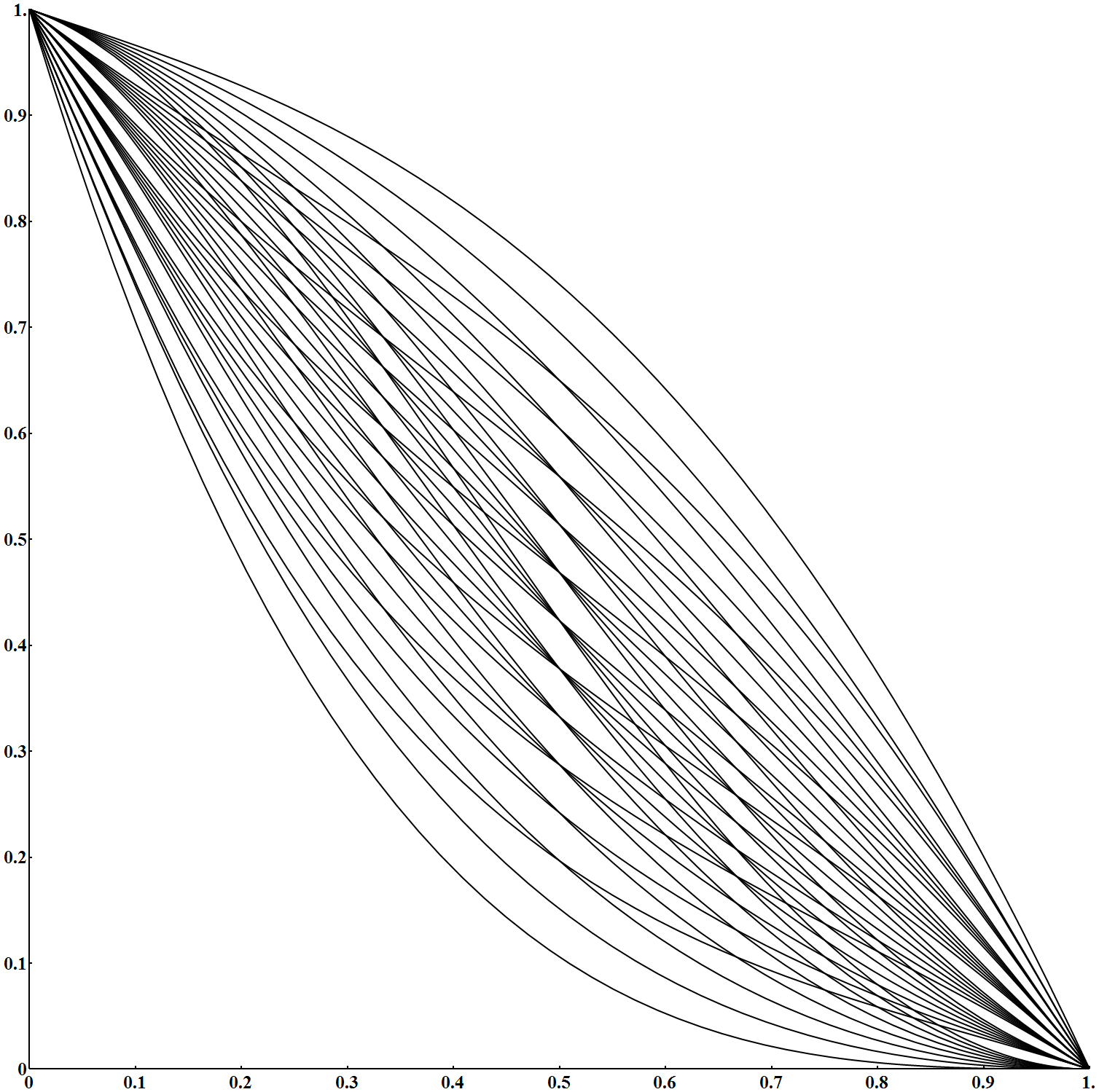} &
    \includegraphics[width = 8 cm, keepaspectratio]{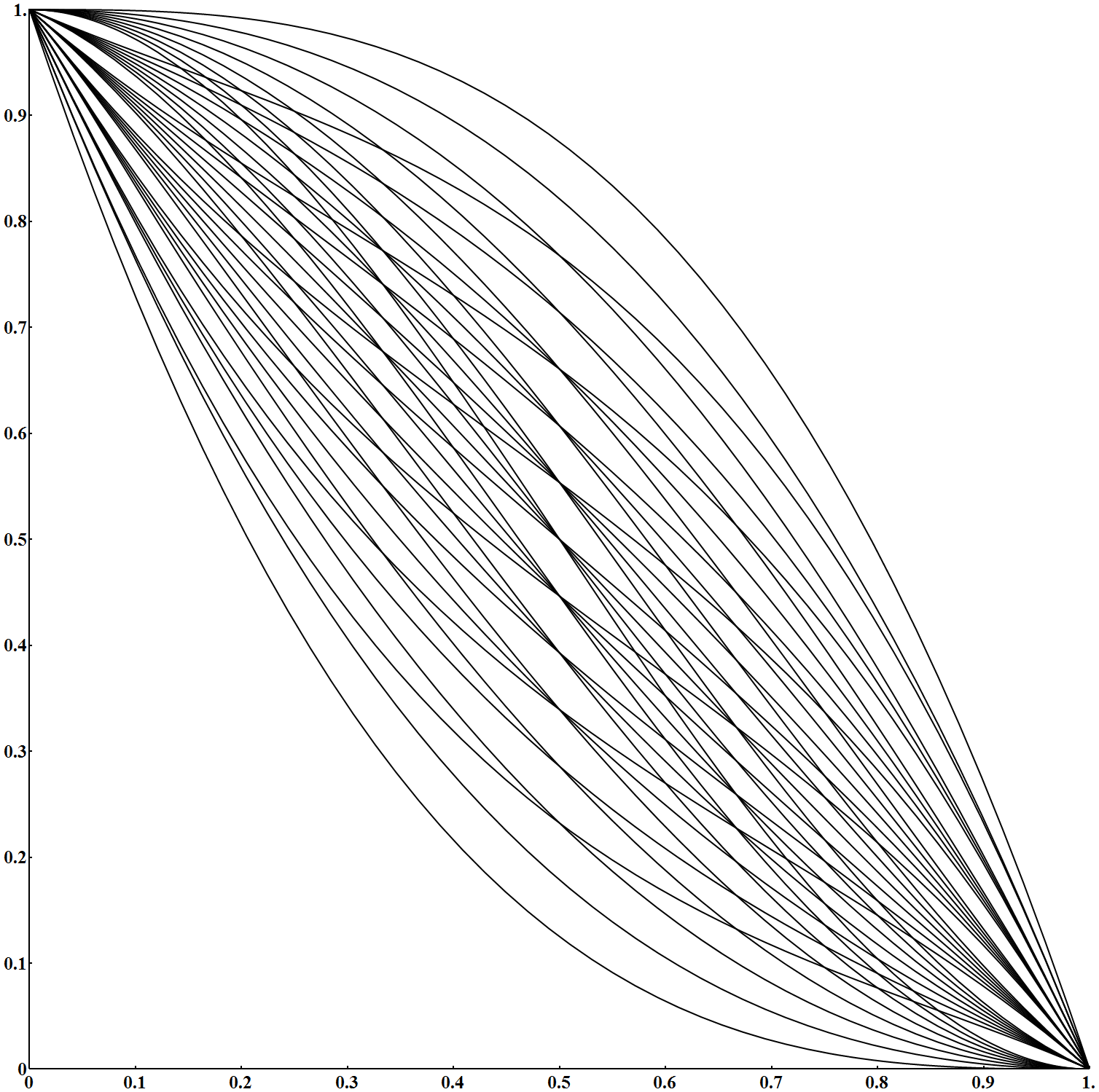} \\
    a) &
    b)
  \end{tabular}
  \caption
  {
    Example of kernels \eqref{eq:KernelsPolynomialExponentThree}.
    $\myfunctionwi{1} = \dfrac{i}{7}$, $\myfunctionwi{2} = \dfrac{j}{7}$, $\myfunctionwi{3} = 1 - \myfunctionwi{1} - \myfunctionwi{2}$, $i = \overline{0..7}$, $j = \overline{0..7 - i}$. a) With exponent. b) Without exponent.
  }
  \label{fig:KernelsPolynomialExponentThree}
\end{figure}

In the case of $m = 3$
\begin{equation}
  \left\{
    \begin{aligned}
      &
      \left(
        \myfunctionwi{1} \left( 1 - h \right)^3
        +
        \myfunctionwi{2} \left( 1 - h \right)^2 \left( 1 + 2 h \right)
        +
        \myfunctionwi{3} \left( 1 - h^3 \right)
      \right)
      \cdot
      \myexp^{-\frac{h}{3}}
      ,
      &
      \text{ if }
      h < 1,
      \\
      &
      0
      ,
      &
      \text{otherwise},
    \end{aligned}
  \right.
  \label{eq:KernelsPolynomialExponentThree}
\end{equation}
where $0 \leq \myfunctionwi{i} \leq 1$, $i = 1, 2, 3$, and $\myfunctionwi{1} + \myfunctionwi{2} + \myfunctionwi{3} = 1$. They are shown in Figure~\ref{fig:KernelsPolynomialExponentThree}a.

From \eqref{eq:RelationshipBetweenKernelsThree}, it follows that $k_1{\left( h \right)}$ has the same form as \eqref{eq:FormFinite}. Notice that $k_3^2{\left( h \right)}$ also has the same form, and the one-dimensional kernel corresponding to $k_3^2{\left( h \right)}$ will be of the same form.

It is possible to extend this approach for anisotropic kernels by adjusting weights and one-dimensional kernel shape in each direction or by stretching the space.

\section{Examples}

\begin{figure} [htb]
  \centering
  \includegraphics[width = 12 cm, keepaspectratio]{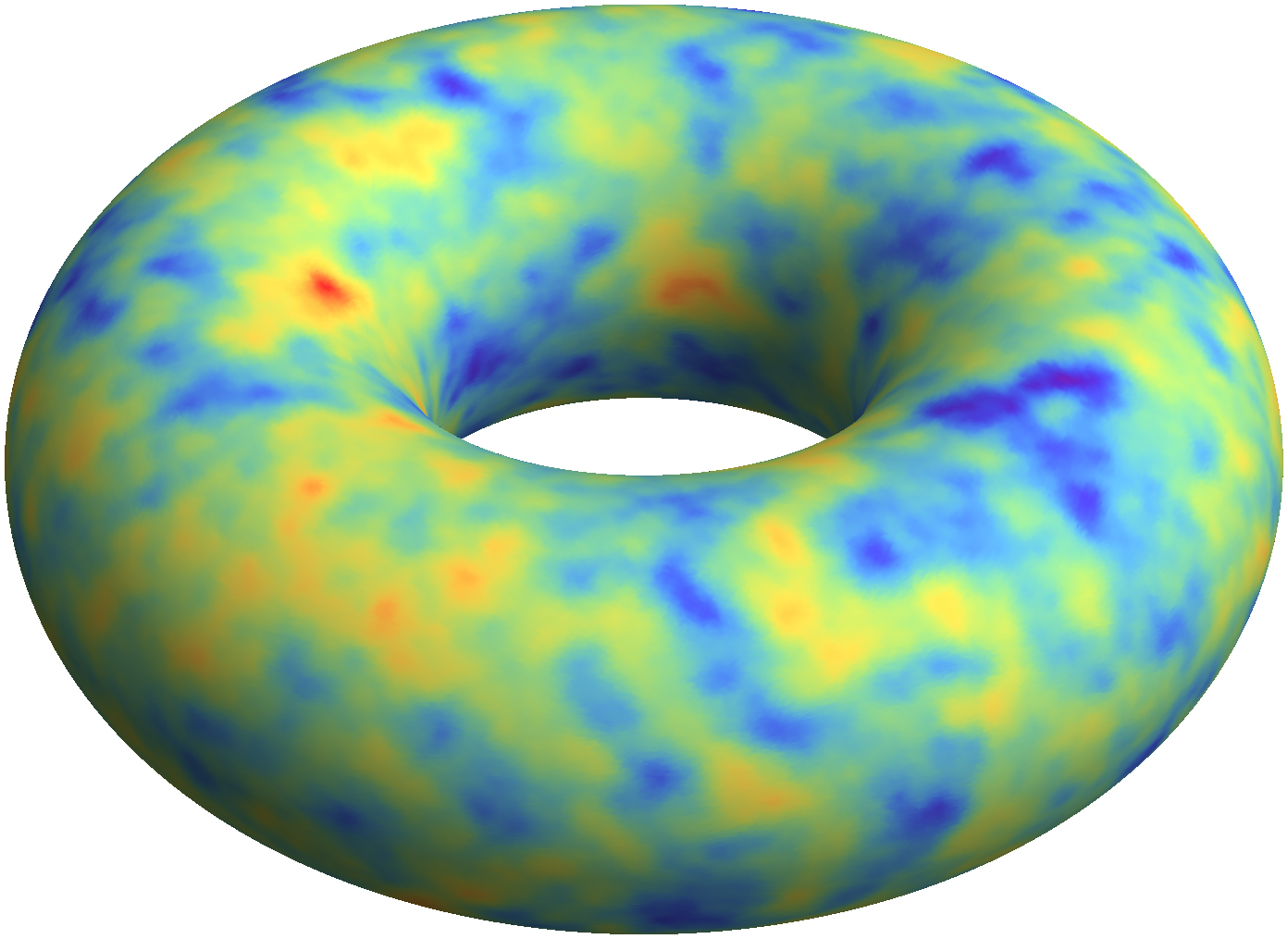} 
  \caption{Unconditional simulation on a torus with inner radius 1, outer radius 3, and kernel $\myexp^{-10 h}$. Number of simulated points $N = 67,108,864$. Number of random directions $N_d = 1,024$.}
  \label{fig:UnconditionalSimulationTorus}
\end{figure}

Figure~\ref{fig:UnconditionalSimulationTorus} shows unconditional simulation on a torus.

\begin{figure} [htb]
  \centering
  \includegraphics[width = \columnwidth, keepaspectratio]{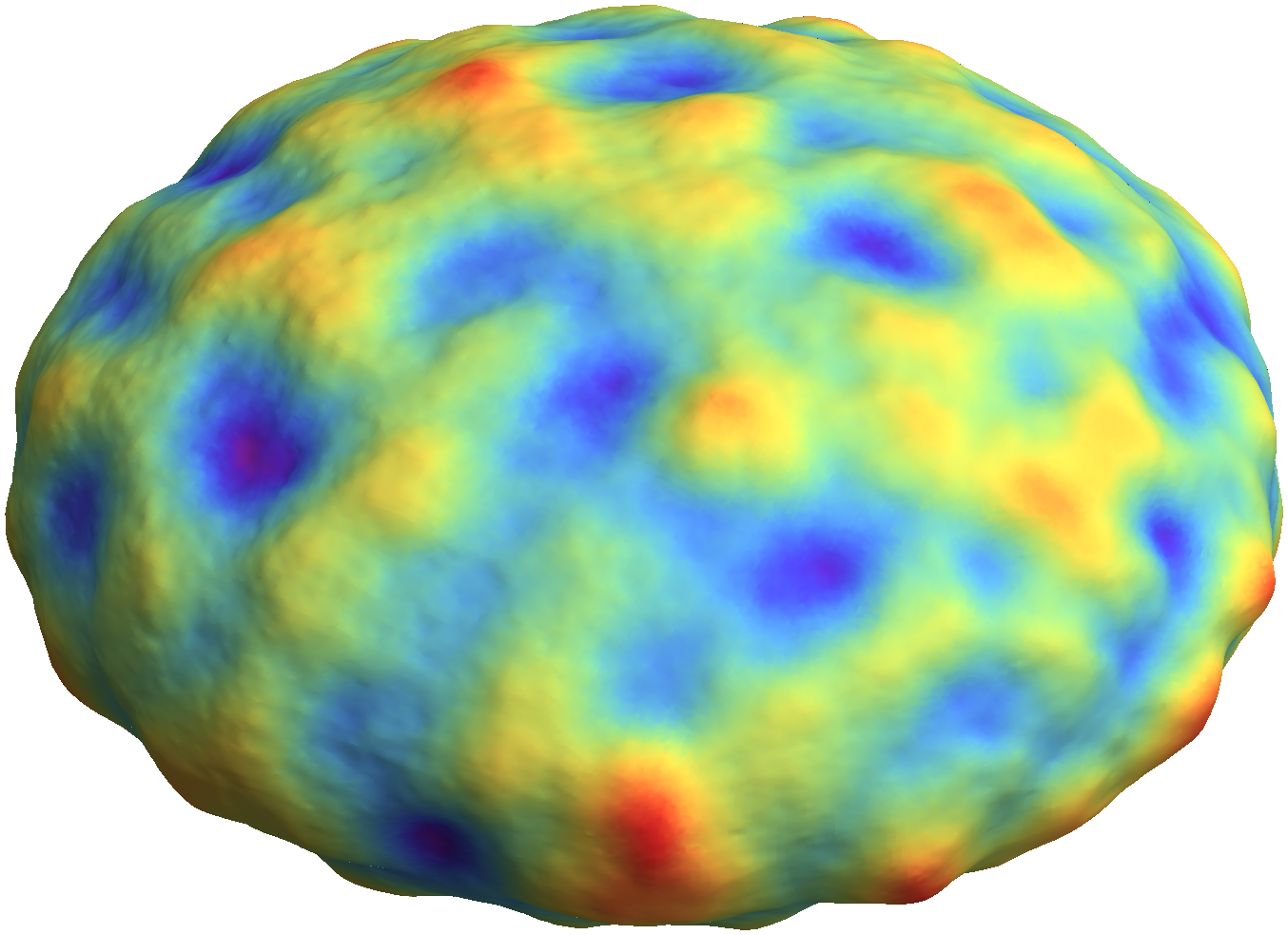}
  \caption
  {
    Unconditional simulation on an oblate spheroid with semimajor axis equals $2$, and semiminor axis equals $1$. The surface is shown as elevation to depict changes in smoothness. The kernel has the form of \eqref{eq:KernelsPolynomialExponentTwo}. It changes smoothly from the kernel on the left
    $
      \left( 1 - \left( 3h \right)^2 \right)
      \cdot
      \myexp^{-h}
    $
    to the kernel on the right
    $
      \left( 1 - 3h \right)^2
      \cdot
      \myexp^{-h}
    $
    with bandwidth equals $\frac{1}{3}$.
    Number of simulated points $N = 262,144$. Number of random directions $N_d = 1,024$.
  }
  \label{fig:UnconditionalSimulatiOnEllipsoid}
\end{figure}

Figure~\ref{fig:UnconditionalSimulatiOnEllipsoid} shows nonstationary unconditional simulation on a spheroid with eccentricity equals $\frac{3}{4}$. The kernel changes smoothly in shape from left to right. $1,024$ unconditional simulations were used to estimate variance and standardize unconditional simulations. Due to changes in the kernel, the right and left parts of the spheroid have different covariance structures. This is visible as a smoother random field on the right compared to the random field on the left.

\section{Convergence Issue}

When the kernel bandwidth tends to be small, the convergence of \eqref{eq:KernelApplied} to the true kernel \eqref{eq:IntegralKernel} becomes slow. The number of required directions makes this method inefficient. This is because the distribution of the number of points that are farther away from the center of the kernel (location $\vec{s}$ in \eqref{eq:KernelApplied}) in three dimensions increases squarely. This is even true for some directions when points are only located on the surface.

\section{Tile Approach}

\begin{figure} [htb]
  \centering
  \begin{tikzpicture}[scale = 2.0]
    \fill [cyan] (2.5, 1) -- (3.5, 1) arc (-90:0:0.5) -- (4, 2.5) arc (0:90:0.5) -- (2.5, 3) arc (90:180:0.5) -- (2, 1.5) arc (180:270:0.5) -- cycle;
    \fill [red] (2.7, 1.8) circle (0.5);
    \tkzInit[xmin = 0, ymin = 0, xmax = 5, ymax = 4]
    \tkzGrid
    \tkzAxeXY
  \end{tikzpicture}
  \caption
  {
    The red circle is the area covered by a finite kernel. The blue rounded square is the tile completely covering the finite kernel.
  }
  \label{fig:OverlappingTiles}
\end{figure}
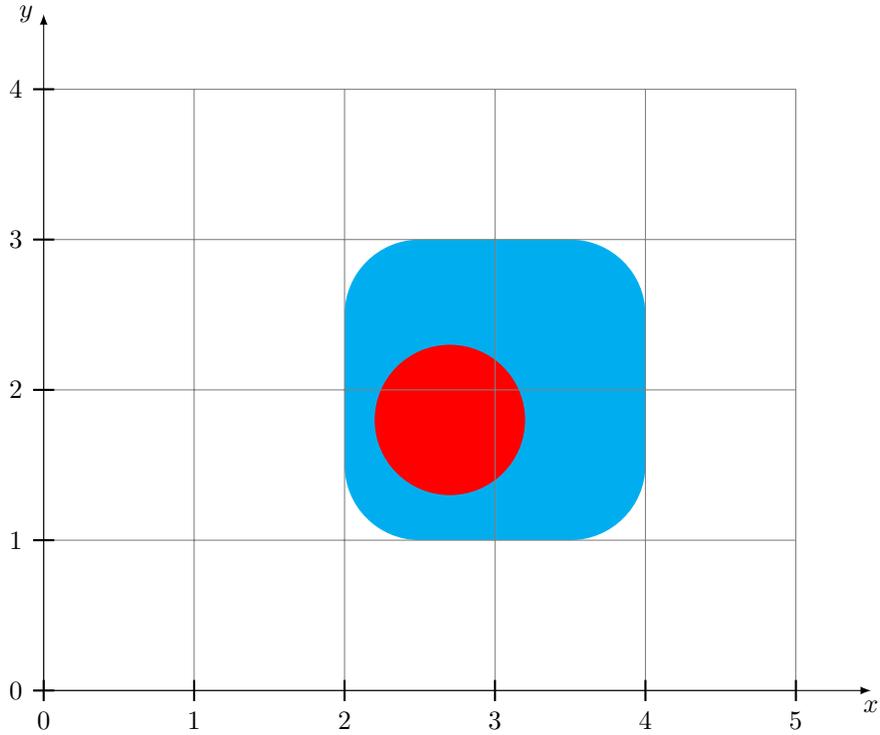

Finite kernel and overlapping tiles overcome the issue described in the previous section. For example, in two dimensions, if all tiles are $2$ by $2$ and start from integer numbers (corners can be rounded), then the kernel with a bandwidth not exceeding $\frac{1}{2}$ will always be completely inside at least one tile; see Figure~\ref{fig:OverlappingTiles}. This is applicable in any number of dimensions.

The bandwidth should not be significantly smaller than the tile width, and the exponential part in the kernel can be dropped. The flexible family of kernels without the exponential part is shown in Figures~\ref{fig:KernelsPolynomialExponent}b and \ref{fig:KernelsPolynomialExponentThree}b.

\begin{figure} [htb]
  \centering
  \begin{tabular}{c c}
    \includegraphics[width = 8.5 cm, keepaspectratio]{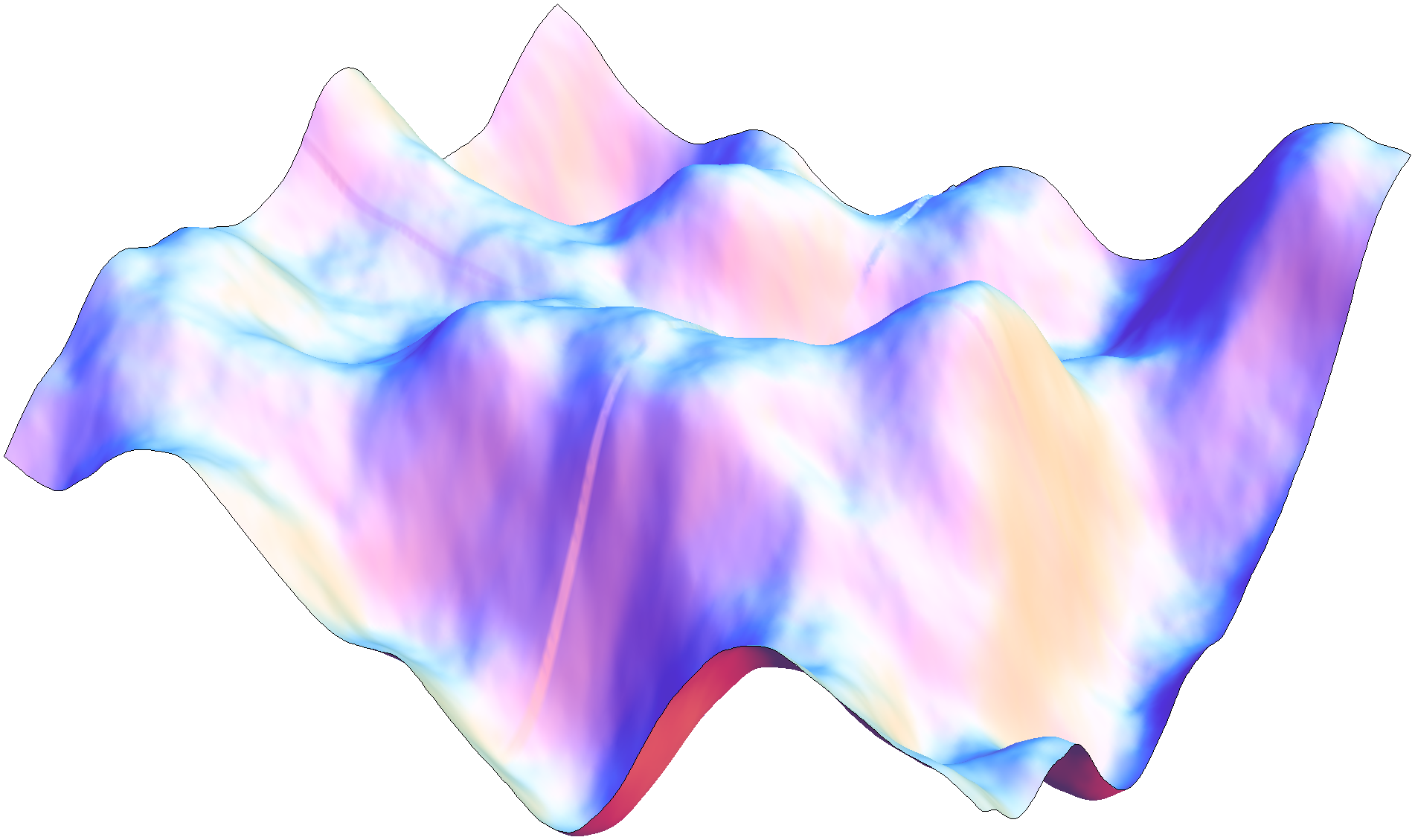} &
    \includegraphics[width = 8.5 cm, keepaspectratio]{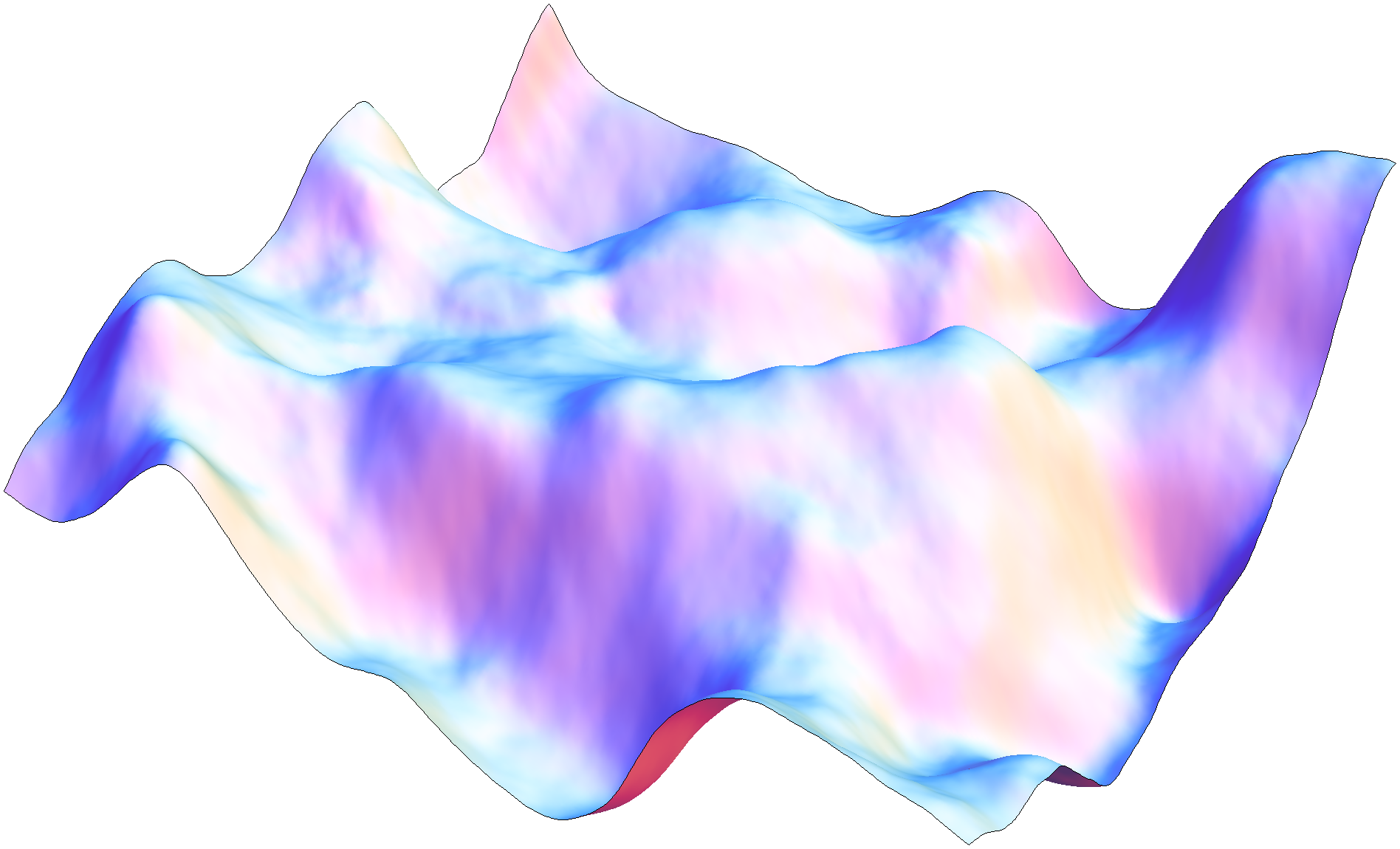} \\
    a) &
    b)
  \end{tabular}
  \caption
  {
    Unconditional simulation in two dimensions by kernel
    $
      k_3{\left( h \right)}
      =
      \left( 1 - 2h \right)^2
    $
    with bandwidth equals $\frac{1}{2}$.
    Number of simulated points $N = 65,536$. a)~The tile approach with the number of random directions $N_d = 1,024$ (uniformly distributed in three dimensions). b)~Direct calculation.
  }
  \label{fig:UnconditionalSimulatiOnPlane}
\end{figure}

Figure~\ref{fig:UnconditionalSimulatiOnPlane}a shows the effect of using the tile approach. The small discontinuities are visible in the horizontal and vertical lines passing through the center. Increasing the number of bands will reduce this effect. Figure~\ref{fig:UnconditionalSimulatiOnPlane}b shows the result of evaluating \eqref{eq:KernelApplied} and \eqref{eq:KernelVariance} directly.

If a continuous result is required, overlapped tiles can be joined smoothly. This will increase calculations by the number of tiles necessary to process. One way to reduce the number of tiles is to use a triangular grid (use nonequilateral triangles for arbitrary surfaces). For each node of the triangular grid, define a tile that covers all neighboring triangles and at least a kernel bandwidth. For each triangle, the barycentric coordinate system will give weights for the tiles. There will only be a maximum of three tiles with nonzero weights; therefore, the complexity is increased three times.

\section{Using Integer Arithmetic}

In the turning band approach, the directions are chosen randomly or pseudorandomly. In this section, another approach will be described to make directions proportional to integer numbers. Using such directions will remove errors from applying fixed steps (bins) in each direction.

\begin{figure} [htb]
  \centering
  \begin{tabular}{c c}
    \includegraphics[width = 8.5 cm, keepaspectratio]{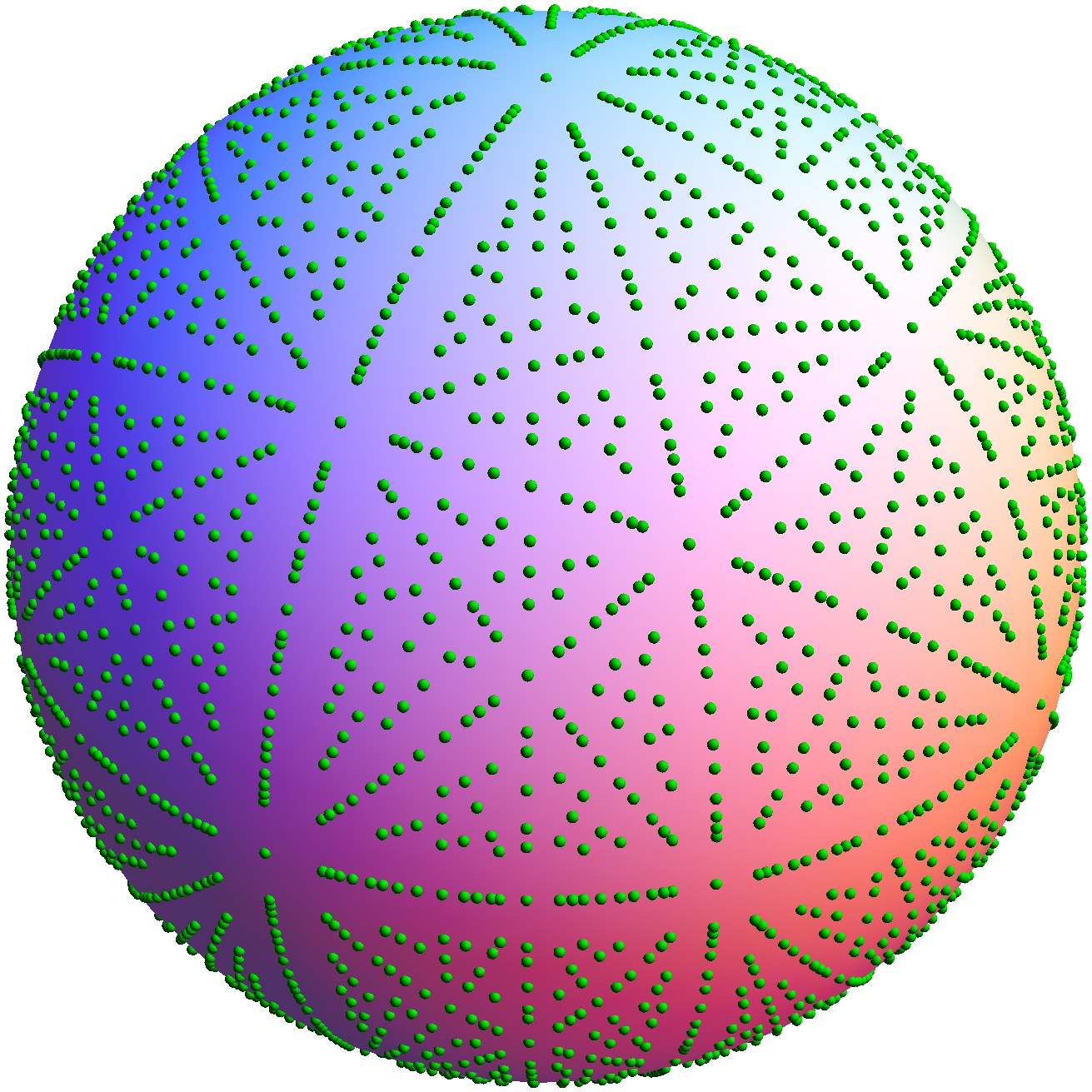} &
    \includegraphics[width = 8.5 cm, keepaspectratio]{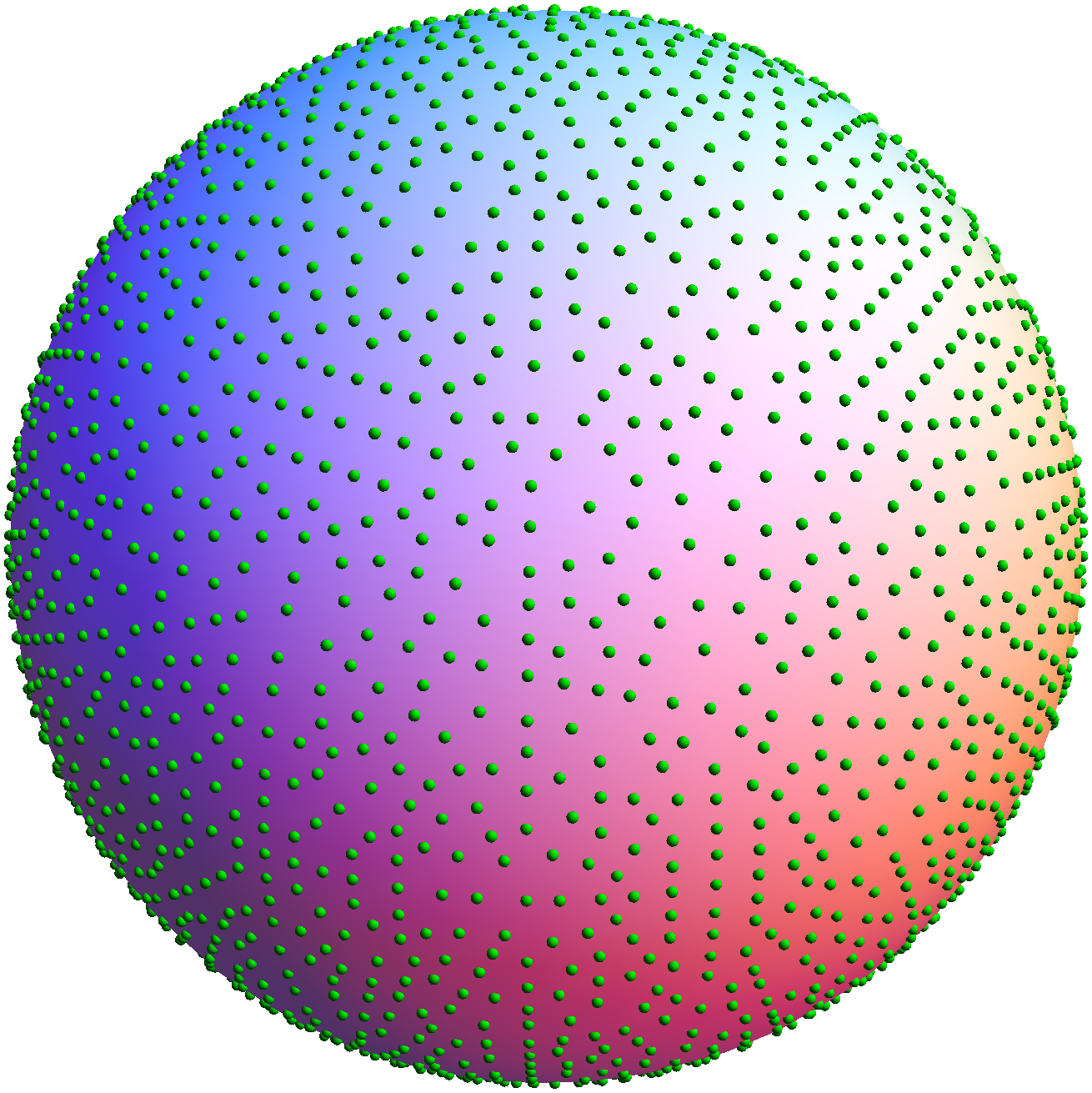} \\
    a) &
    b)
  \end{tabular}
  \caption
  {
    a) All integer directions are shown as dots on the sphere for $S = 100$. The number of unique directions equals $1,729$ (counting $\vec{v}$ and $-\vec{v}$ once). b) A uniform set of integer directions satisfying that any two directions are at least $3 \degree$ apart. The number of directions equals $1,405$.
  }
  \label{fig:IntegerDirections}
\end{figure}

Let's take a sphere of radius $R = \sqrt{S}$, where $S$ is a natural number. If the radius $R$ approaches infinity, all integer vectors inside the sphere (all integer vertices except the center of a coordinate system) tend to be uniformly distributed. However, the unique directions are not uniformly distributed (see example in Figure~\ref{fig:IntegerDirections}a). But assigning a proper weight to each direction will resolve it. The appropriate weights can be assigned as proportionate to the number of vectors with the same direction or to the area of the spherical Voronoi diagram cells.

Selecting only directions separated by some degree will make a uniform coverage (see Figure~\ref{fig:IntegerDirections}b). The list of integer directions is shown in the \hyperref[SectionAppendix]{appendix}. This is similar to using a set of pseudorandom directions as described in \cite{RandomDirections} and \cite{TurningBandSimulationsInThreeDimensions}.

Because each direction is represented as $\alpha \cdot \left( x, y, z \right)$, where $\alpha$ is a real number (square root of a natural number) and $x$, $y$, and $z$ are integers, any coordinate with integer components will be projected to a position $\alpha \cdot i$, where $i$ is an integer number.

If random points are placed on an integer grid, there are no errors due to binning. If the resultant location also has an integer coordinate, there is no additional error. Notice that the kernel \eqref{eq:Form} or \eqref{eq:FormFinite} can be calculated for any position on direction, see \cite{Smoothing} and \cite{SmoothingNetwork}. The turning band approach \cite{TwoDimensionalSimulationByTurningBands}, \cite{TurningBandTwoDimensional}, \cite{ClosedFormSolutionsOfTheTwoDimensionalTurningBandEquation}, and \cite{TurningBandSimulationsInThreeDimensions} might also benefit from using integer arithmetic.

\section{Conclusion}

The new approach to producing unconditional nonstationary simulations on arbitrary surfaces is equivalent to the kernel convolution approach with the kernel shape depending on location (\cite{KernelConvolution1996}, \cite{KernelConvolution1998}, and \cite{KernelConvolutionFFT}). Applying a kernel to the arbitrary surfaces produces a larger class of valid covariance functions than the class of valid covariance functions in three-dimensional space.

Precalculating projections of random points in all directions for all tiles might be necessary for the efficiency of the approach. Nevertheless, even with precalculations, this approach is not efficient. The better approach is based on kernel convolution using FFT as described in \cite{ConvolutionFFT}.

New flexible classes of kernels constructed from polynomials are described in this paper. Because they are linear combinations of some basis kernels, constructing covariances for all pairs of basis kernels is sufficient to calculate covariance between any two kernels. Unconditional simulation for a nonstationary kernel is a weighted sum of unconditional simulations corresponding to basis kernels for the same set of random points.

Using integer arithmetic, an error introduced in the turning band approach due to binning along each direction can be avoided.

\section{Acknowledgment}

The author would like to thank David Burrows for help with area preserving transformation from authalic sphere to spheroid (authalic sphere is a sphere were the surface area equals the spheroid surface area). This transformation was used in simulating uniformly distributed random points on the surface of a spheroid for the creation of Figure~\ref{fig:UnconditionalSimulatiOnEllipsoid}.


\newcommand{\doi}[1]{\textsc{doi}: \href{http://dx.doi.org/#1}{\nolinkurl{#1}}}


\begingroup
\raggedright
\bibliographystyle{IEEEtran}
\bibliography{ATurningBandApproachToKernelConvolutionForArbitrarySurfaces}
\endgroup



\section*
{
  Appendix. List of Integer Directions
  \label{SectionAppendix}
}

Integer directions that are at least $3 \degree$ apart:
$ \left( 1, 0, 0 \right) $,
$ \frac{\left( 1, 1, 0 \right)}{ \sqrt{2}} $,
$ \frac{\left( 1, 1, 1 \right)}{ \sqrt{3}} $,
$ \frac{\left( 2, 1, 0 \right)}{ \sqrt{5}} $,
$ \frac{\left( 2, 1, 1 \right)}{ \sqrt{6}} $,
$ \frac{\left( 3, 1, 0 \right)}{ \sqrt{10}} $,
$ \frac{\left( 2, 2, 1 \right)}{3} $,
$ \frac{\left( 3, 1, 1 \right)}{ \sqrt{11}} $,
$ \frac{\left( 3, 2, 0 \right)}{ \sqrt{13}} $,
$ \frac{\left( 4, 1, 0 \right)}{ \sqrt{17}} $,
$ \frac{\left( 3, 2, 1 \right)}{ \sqrt{14}} $,
$ \frac{\left( 4, 1, 1 \right)}{ \sqrt{18}} $,
$ \frac{\left( 3, 2, 2 \right)}{ \sqrt{17}} $,
$ \frac{\left( 3, 3, 1 \right)}{ \sqrt{19}} $,
$ \frac{\left( 4, 2, 1 \right)}{ \sqrt{21}} $,
$ \frac{\left( 4, 3, 0 \right)}{5} $,
$ \frac{\left( 5, 1, 1 \right)}{ \sqrt{27}} $,
$ \frac{\left( 5, 2, 0 \right)}{ \sqrt{29}} $,
$ \frac{\left( 6, 1, 0 \right)}{ \sqrt{37}} $,
$ \frac{\left( 3, 3, 2 \right)}{ \sqrt{22}} $,
$ \frac{\left( 4, 3, 1 \right)}{ \sqrt{26}} $,
$ \frac{\left( 5, 2, 1 \right)}{ \sqrt{30}} $,
$ \frac{\left( 4, 3, 2 \right)}{ \sqrt{29}} $,
$ \frac{\left( 4, 4, 1 \right)}{ \sqrt{33}} $,
$ \frac{\left( 5, 2, 2 \right)}{ \sqrt{33}} $,
$ \frac{\left( 5, 3, 1 \right)}{ \sqrt{35}} $,
$ \frac{\left( 6, 2, 1 \right)}{ \sqrt{41}} $,
$ \frac{\left( 7, 1, 1 \right)}{ \sqrt{51}} $,
$ \frac{\left( 4, 3, 3 \right)}{ \sqrt{34}} $,
$ \frac{\left( 5, 3, 2 \right)}{ \sqrt{38}} $,
$ \frac{\left( 5, 4, 1 \right)}{ \sqrt{42}} $,
$ \frac{\left( 6, 3, 1 \right)}{ \sqrt{46}} $,
$ \frac{\left( 9, 1, 0 \right)}{ \sqrt{82}} $,
$ \frac{\left( 5, 4, 2 \right)}{ \sqrt{45}} $,
$ \frac{\left( 6, 3, 2 \right)}{7} $,
$ \frac{\left( 6, 4, 1 \right)}{ \sqrt{53}} $,
$ \frac{\left( 7, 3, 1 \right)}{ \sqrt{59}} $,
$ \frac{\left( 7, 4, 0 \right)}{ \sqrt{65}} $,
$ \frac{\left( 8, 2, 1 \right)}{ \sqrt{69}} $,
$ \frac{\left( 5, 4, 3 \right)}{ \sqrt{50}} $,
$ \frac{\left( 7, 3, 2 \right)}{ \sqrt{62}} $,
$ \frac{\left( 10, 1, 1 \right)}{ \sqrt{102}} $,
$ \frac{\left( 6, 4, 3 \right)}{ \sqrt{61}} $,
$ \frac{\left( 6, 5, 2 \right)}{ \sqrt{65}} $,
$ \frac{\left( 6, 6, 1 \right)}{ \sqrt{73}} $,
$ \frac{\left( 7, 4, 2 \right)}{ \sqrt{69}} $,
$ \frac{\left( 7, 6, 0 \right)}{ \sqrt{85}} $,
$ \frac{\left( 10, 2, 1 \right)}{ \sqrt{105}} $,
$ \frac{\left( 5, 5, 4 \right)}{ \sqrt{66}} $,
$ \frac{\left( 7, 6, 1 \right)}{ \sqrt{86}} $,
$ \frac{\left( 10, 3, 1 \right)}{ \sqrt{110}} $,
$ \frac{\left( 7, 4, 4 \right)}{9} $,
$ \frac{\left( 8, 6, 1 \right)}{ \sqrt{101}} $,
$ \frac{\left( 9, 5, 1 \right)}{ \sqrt{107}} $,
$ \frac{\left( 11, 3, 2 \right)}{ \sqrt{134}} $,
$ \frac{\left( 11, 4, 1 \right)}{ \sqrt{138}} $,
$ \frac{\left( 8, 7, 2 \right)}{ \sqrt{117}} $,
$ \frac{\left( 11, 5, 1 \right)}{ \sqrt{147}} $,
$ \frac{\left( 7, 6, 5 \right)}{ \sqrt{110}} $,
$ \frac{\left( 11, 4, 3 \right)}{ \sqrt{146}} $,
$ \frac{\left( 7, 6, 6 \right)}{11} $,
$ \frac{\left( 8, 7, 4 \right)}{ \sqrt{129}} $,
$ \frac{\left( 10, 5, 4 \right)}{ \sqrt{141}} $,
$ \frac{\left( 11, 7, 1 \right)}{ \sqrt{171}} $,
$ \frac{\left( 17, 1, 1 \right)}{ \sqrt{291}} $,
$ \frac{\left( 18, 1, 0 \right)}{ \sqrt{325}} $,
$ \frac{\left( 13, 4, 3 \right)}{ \sqrt{194}} $,
$ \frac{\left( 16, 4, 1 \right)}{ \sqrt{273}} $,
$ \frac{\left( 18, 3, 1 \right)}{ \sqrt{334}} $,
$ \frac{\left( 10, 7, 6 \right)}{ \sqrt{185}} $,
$ \frac{\left( 10, 9, 4 \right)}{ \sqrt{197}} $,
$ \frac{\left( 11, 11, 1 \right)}{ \sqrt{243}} $,
$ \frac{\left( 12, 8, 3 \right)}{ \sqrt{217}} $,
$ \frac{\left( 12, 7, 6 \right)}{ \sqrt{229}} $,
$ \frac{\left( 14, 12, 1 \right)}{ \sqrt{341}} $,
and
$ \frac{\left( 17, 12, 1 \right)}{ \sqrt{434}} $.
The complete list of directions is formed by considering for each vector all signs and permutations of its components.

\end{document}